\documentstyle[12pt]{article}

\def\Dirac{{\raise0.09em\hbox{/}}\kern-0.69em D}

\def\bbbc{{\mathchoice
{\setbox0=\hbox {$\displaystyle\rm C$}\hbox
{\hbox to0pt{\kern0.4\wd0\vrule height0.9\ht0\hss}\box0}}
{\setbox0=\hbox {$\textstyle\rm C$}\hbox
{\hbox to0pt{\kern0.4\wd0\vrule height0.9\ht0\hss}\box0}}
{\setbox0=\hbox {$\scriptstyle\rm C$}\hbox
{\hbox to0pt{\kern0.4\wd0\vrule height0.9\ht0\hss}\box0}}
{\setbox0=\hbox {$\scriptscriptstyle\rm C$}\hbox
{\hbox to0pt{\kern0.4\wd0\vrule height0.9\ht0\hss}\box0}}
}}

\newcommand{\AmS}{{\protect\the\textfont2
  A\kern-.1667em\lower.5ex\hbox{M}\kern-.125emS}}

\title{Classical Gravity on Fuzzy Space-Time}

\author{J. Madore \\
        Laboratoire de Physique Th\'eorique et Hautes
        Energies\thanks{Laboratoire associ\'e au CNRS, URA D0063}\\
        Universit\'e de Paris-Sud, B\^at. 211, F-91405 Orsay\\
       }

\begin{document}

\maketitle

\begin{abstract}
A review is made of recent efforts to find relations between the
commutation relations which define a noncommutative geometry and the
gravitational field which remains as a shadow in the commutative limit.
\end{abstract}


\section{Historical Introduction}

The position $x$ and the momentum $p$ of a classical particle can be
simultaneously measured and $(x,p)$ defines a point in classical
phase-space. The set of polynomials in the variables $(x, p)$ can be
added and multiplied; they form a commutative algebra. When the particle
is quantized the points disappear; because of the Heisenberg uncertainty
relations the two operators $\hat x$ and $\hat p$ can no longer be
simultaneously measured.  However, it still makes sense to consider the
algebra of polynomials in them.  It is a noncommutative algebra which
Dirac \cite{Dirac} referred to as the `quantum algebra'.

A vector in classical phase space can be naturally identified with a
derivation of the algebra of polynomials and as such can be generalized
to the quantized case. Dirac called it a `quantum differentiation'. From
the correspondence principle we see that
$$
\lim_{\hbar \rightarrow 0} {i \over \hbar} [\hat p, \hat f] 
\rightarrow {\partial f \over \partial x}, \quad
\lim_{\hbar \rightarrow 0} {i \over \hbar} [\hat x, \hat f] 
\rightarrow - {\partial f \over \partial p}.
$$
The $\hat f$ is here an arbitrary element of the `quantum algebra'.
The quantized version of the vector $X = (X_x, X_p)$ is then the 
derivation of the algebra given by 
$$
\hat Xf = {i \over \hbar} \hat X_x [\hat p, \hat f] - 
{i \over \hbar} \hat X_p [\hat x, \hat f].
$$
Because of the ordering problem we see that a `quantum derivation'
defines a unique vector but that the inverse is not true.

Some decades later von Neumann introduced the term `noncommutative
geometry' to refer in general to a geometry in which an algebra of
functions is replaced by a noncommutative algebra. As in the phase-space
example coordinates are replaced by generators of the algebra. Since
these do not commute they cannot be simultaneously diagonalized and the
space disappears. We shall argue with an example below that, just as
Bohr cells replace classical-phase-space points, the appropriate
intuitive notion to replace a `point' is a `cell'.  

But `geometry' is more than a set of points and so more is needed that
just an algebra.  This problem was solved by Connes \cite{Connesa,Connesb} 
who introduced the notion of `noncommutative differential geometry' to
refer in general to a noncommutative geometry with an associated
differential calculus.  Just as it is possible to give many
differential structures to a given topological space it is possible to
define many differential calculi over a given algebra. We shall use the
term `noncommutative geometry' to mean `noncommutative differential
geometry' in the sense of Connes.

In order to cope with the divergences in the newly-discovered quantum
field theory, throughout the 1930's Heisenberg and others flirted with
the idea of replacing space-time with a fundamental lattice. The lattice
spacing would serve as cut-off. The difficulty here is that the lattice
destroys Lorentz invariance. A solution to this problem was proposed
some time later by Snyder \cite{Snyder} who showed that by using
noncommutative `coordinates' it was possible to have a version of
`space-time' which had some of the desirable features of a lattice but
which was nevertheless Lorentz invariant. 

When referring to the version of space-time which we shall describe here
we use the adjective `fuzzy' to underline the fact that points are
ill-defined \cite{Madorea,Madoreb}.  Since the algebraic structure is
described by commutation relations the qualifier `quantum' was used by
Snyder and by others \cite{DFR,MadoreMouradc}.  This latter expression
is unfortunate since the structure has no immediate relation to quantum
mechanics and also it leads to confusion with a `space-time' on which a
`quantum' Lorentz group acts \cite{Castellani}.

It is worth mentioning something which we do {\em not} mean by the
expression `fuzzy space-time'. To explain the Zitterbewegung of an
electron Schr\"odinger and others considered center-of-mass position
operators of the form
$$
q^i = \hat x^i + m^{-2} (\hat S \times \hat p)^i.
$$
Here $\hat S$ is the spin polarization vector.  Because of the canonical
commutation relations the operators $q^i$ do not commute. One could
introduce an algebra generated by them, for each value of the time
coordinate, and even a differential calculus to construct a
noncommutative geometry. The algebra of the commutative limit would be
however simply the real (or complex) numbers, the possible values of a
function at the position of the particle; it would not be an algebra of
functions on space-time. An associated noncommutative geometry would be
a noncommutative generalization of a point. This example has a membrane
generalization.  The fact that the target-space coordinates of an
extended object do not commute does not mean that the noncommutative
algebra which they generate is a noncommutative version of space-time.
It could however be considered as a noncommutative Kaluza-Klein
extension of a point of ordinary space-time. For example if one truncates
\cite{deWitLuscherNicolai} the modes of a membrane and modifies the
product to construct a matrix algebra the resulting geometry
\cite{DuboisVioletteKernerMadore} is a matrix extension of space-time.
Conversely a static `membrane' in a fuzzy space-time as we define it can
only have a finite number of modes and will be described by some
finite-dimensional quotient algebra.
\vfill

\section{Fuzzy Space-Time}

By a `fuzzy' space-time we mean an algebra with 4 generators $q^\mu$
which do not commute:
\begin{equation}
[q^\mu, q^\nu] = i \mu_P^{-2} q^{\mu\nu}               \label{comrel}
\end{equation}
In the limit where the Planck mass $\mu_P$ tends to infinity we wish to
identify the generators with the coordinates of ordinary Minkowski
space-time:
\begin{equation}
\lim_{\mu_P \rightarrow \infty} q^\mu = x^\mu.
\end{equation}
The structure of the algebra is given, for example, by the commutation
relations $[q^\lambda, q^{\mu\nu}]$. Several possibilities have been
considered \cite{Snyder,Madoreb,DFR}.

There are several reasons for which one might be lead to consider fuzzy
space-times. Effectively when one computes for example Feynman diagrams
one is forced to introduce a cut-off $\Lambda$, which can certainly be
taken to be less than the Planck mass. This introduces a fuzziness in
the definition of space-time points of the order of at least
$\mu_P^{-1}$.  Since this is an unavoidable feature of particle physics
one might as well include it as part of the geometry of space-time. It
can be argued \cite{MadoreMourada} that noncommuting coordinates are a
natural consequence of the quantum fluctuations of the gravitational
field. We would like to argue in fact that the commutation relations of
the algebra determine and are determined by the gravitational field. The
gravitational field can be then considered to regularize the ultraviolet
divergences through the introduction of a noncommutative structure on
space-time.  We have already mentioned the motivation of Snyder to
introduce a Lorentz-invariant version of a lattice.  The most compelling
argument however in favour of studying fuzzy versions of space-time is
that they constitute a natural generalization of ordinary space-time and
they can be studied. It is possible to generalize to the noncommutative
case the notions of vectors, covectors, metrics, connections etc.

\subsection{Problems}

A satisfactory definition of the noncommutative extension of Riemannian
curvature has not yet been found \cite{CuntzQuillen,DuboisVioletteM3a}.
Invariants are another problem. Cyclic cohomology has been used
\cite{Connesa,Connesb} to define topological invariants but the ordinary
invariants which might lead to an action principle have not been found.
Indeed we are not in a position to argue that there is even a valid
action principle. A discussion of this point has been made by Connes and
co-workers in a series of articles
\cite{KalauWalze,AckermannTolksdorf,Connesc,ChamseddineConnes} but the
definition which these authors propose is valid only on the
noncommutative generalizations of compact spaces with
euclidean-signature metrics.

\subsection{General relations}

Consider the algebra generated by the `coordinates' $q^\mu$ and the
following diagram
\begin{equation}
\def\normalbaselines{\baselineskip=18pt}
\begin{array}{ccc}
\mbox{Algebra}&\Longleftarrow&\mbox{Calculus}\\ 
\Downarrow && \phantom{(?)} \Downarrow \; \Uparrow \, (?) \\
\mbox{Cut-off} &&\hbox{Gravitation}
\end{array} 
\def\normalbaselines{\baselineskip=12pt}
\end{equation}
The top arrow is a mathematical triviality. We shall define it in the
next section when we recall the definition of a differential calculus.
The left arrow is the `pointlessness' property of noncommutative
geometry. Since points have been eliminated there can be no 
divergences \cite{Snyder,Madoreb,DFR}.  As mentioned above this is one
of the principal motivations of studying noncommutative geometry. The
relations expressed by the right arrow are new results
\cite{DimakisMadore,MadoreMouradc}. We shall argue below using simple
models that the gravitational field is determined by the differential
calculus. We would like to argue that the inverse is also true but this
is less certain. If we could argue that there is a `natural'
differential calculus over every `natural' noncommutative generalization
of space-time then we could conclude that the commutation relations of
the algebra determine and are determined by the gravitational field. In
any case to within the uncertainty expressed by the question mark the
diagram expresses a possible realization of the old idea of Pauli
\cite{Deser,IshamSalamStrathdee} that the gravitational field can be
used to regularize ultraviolet divergences.
\vfill

\section{Differential Calculi}

We recall that a differential calculus 
over an algebra ${\cal A}$ is another (graded) algebra 
\begin{equation}
\Omega^*({\cal A}) = \bigcup_p \Omega^p({\cal A})
\end{equation}
which gives a differential structure to ${\cal A}$. The elements of
$\Omega^p({\cal A})$ are called $p$-forms. There is a linear map $d$
which takes $p$-forms into $(p+1)$-forms and which satisfies a graded
Leibniz rule as well as the condition $d^2=0$.  By definition
$\Omega^0({\cal A}) = {\cal A}$. In general there can be several
differential calculi over a given algebra.  For a more detailed
discussion within the context of noncommutative geometry we refer to the
primary sources \cite{Connesa,Connesb} or to secondary sources, for
example, \cite{Madoreb,MadoreMouradb}. 

In ordinary geometry the 1-forms $\Omega^1({\cal A})$ are dual to the
vector fields. There is also an interesting relation between the
differential $d$ and the Dirac operator $\Dirac$ which we mention.  Let
$\psi$ be a Dirac spinor and $f$ a smooth function. It is
straightforward to see that
\begin{equation}
\partial_\lambda f \gamma^\lambda\psi = - [i\Dirac, f] \psi.
\end{equation}
If we make the replacement $\gamma^\lambda \mapsto dx^\lambda$ the left-hand
side becomes equal to $df \psi$ and we can write the differential as a
commutator: 
\begin{equation}
df = - [i\Dirac, f].
\end{equation} 
It would be natural to try to generalize this relation to higher-order
forms by using a graded commutator on the right-hand side.
Because $dx^\mu dx^\nu + dx^\nu dx^\mu = 0$ whereas
$\gamma^\mu \gamma^\nu + \gamma^\nu \gamma^\mu \neq 0$ one would find
that $d^2 \neq 0$. We shall encounder a similar problem in a model in
the next section.

\section{Some Simple Modles}

The simplest examples of noncommutative algebras are furnished by matrix
algebras. A less trivial example is furnished by the quantum plane.

\subsection{The Connes-Lott models}

Consider the algebra $M_2(\bbbc)$ of $2 \times 2$ complex
matrices which we write as $M_2(\bbbc) = M^+_2 \oplus M^-_2$ where
$M^+_2$ is the algebra of diagonal matrices and $M^-_2$ is the set of
off-diagonal matrices.  Set ${\cal A} = M^+_2$ and define
\cite{ConnesLott} a differential calculus over ${\cal A}$ by
\begin{equation}
\Omega^{2p}({\cal A}) = M^+_2, \qquad
\Omega^{2p+1}({\cal A}) = M^-_2
\end{equation}
for all $p \geq 0$. Let $\eta$ be any anti-hermitian element of $M^-_2$.
One can define a differential by
\begin{equation}
d\alpha = - [\eta, \alpha]
\end{equation}
where the commutator is a graded commutator. The $\eta$ is a generalized
Dirac operator. If one chooses $\eta^2 = -1$ it follows that $d^2 = 0$.
One can think of ${\cal A}$ as the algebra of functions on 2 points and
the differential as the finite-difference operator.

A slightly less trivial example is furnished by the algebra $M_3(\bbbc)$
of $3 \times 3$ complex matrices which we write as 
$M_3(\bbbc) = M^+_3 \oplus M^-_3$ where $M^+_3 = M_2(\bbbc)\times \bbbc$
is the algebra of block-diagonal matrices and $M^-_3$ is the remainder.
Set ${\cal A} = M^+_3$ and define \cite{ConnesLott} a differential
calculus over ${\cal A}$ by
$$
\Omega^0({\cal A}) = M^+_3, \quad
\Omega^1({\cal A}) = M^-_3, \quad
\Omega^2({\cal A}) = \bbbc
$$
and $\Omega^p({\cal A}) = 0$, $p \geq 3$.  Let $\eta$ be any
anti-hermitian element of $M^-_3$.  One can define again a differential
as above. It is not possible however to impose the condition $\eta^2 = -1$
and it was necessary to define the 2-forms as the projection of $M^+_3$
onto the single factor $\bbbc$ in order to have the relation $d^2 = 0$.
The problem here is similar to that in the ordinary case where also 
$\Dirac^2 \neq 1$.  One can think of ${\cal A}$ again as an algebra of
functions on two points but with an additional algebraic structure on one
of the points. The differential can no longer be considered as a
finite-difference operator.

\subsection{Derivation-based models}

Consider ${\cal A} = M_2(\bbbc)$ and let $\Omega^*({\cal A})$ be the
differential calculus \cite{DuboisViolette,DuboisVioletteKernerMadore}
based on the Lie algebra of all derivations of ${\cal A}$.  In this case
$\Omega^1({\cal A})$ has an anti-commuting basis of 3 elements
$\theta^i$ which commute with the elements of the algebra:
\begin{equation}
f \theta^i = \theta^i f, \qquad \theta^i \theta^j = - \theta^j \theta^i.
\end{equation}
An arbitrary element $\xi$ of $\Omega^1({\cal A})$ can be written in the
form $\xi = \xi_i \theta^i$ where the $\xi_i$ are elements of ${\cal A}$.
Let $\lambda_i$ be the (suitably normalized) Pauli matrices. The special
1-form $\theta = - \lambda_i \theta^i$ is a generalized Dirac operator.
The first Connes-Lott model is a singular contraction of this one
\cite{MadoreMouradSitarz}.

The algebra $M_2(\bbbc)$ can be used to furnish a very simple example of
what we meant in the Introduction by an `invariant version of a
lattice'. Consider the ordinary round 2-sphere in 3 dimensions and as an
extreme lattice approximation consider the north and south poles.
Functions on these two points can be put in one-to-one correspondence
with diagonal $2 \times 2$ matrices.  The lattice approximation has as
algebra of functions a (commutative) 2-dimensional algebra.  

There is however a more interesting approximation which consists in
considering the 2-sphere as a classical spin and approximating it by a
spin 1/2. The corresponding algebra of observables is the algebra of all
$2 \times 2$ matrices. In general two observables cannot be
simultaneously measured but any single observable has 2 eigenvalues
which correspond to its value on the two `points'. The lattice algebra
is obviously not invariant under the adjoint action of the rotation
group but the full matrix algebra is invariant.  

The algebra ${\cal A} = M_n(\bbbc)$ of $n \times n$ complex matrices can
be used to construct an $SO_3$-invariant `lattice' approximation to the
sphere $S^2$ with `lattice spacing' proportional to $1/n$. There are no
points but the sphere is divided into $n$ cells.  This `fuzzy sphere'
has been studied from several points of view
\cite{Madorea,Madoreb,Madorec,GM,GP,GKPa} and a super-symmetric version
has been proposed \cite{GKPb}.

\subsection{The quantum plane}

The quantum plane is the algebra ${\cal A}$ generated by `variables' 
$x$ and $y$ which satisfy the relation
$$
xy = qyx,
$$
where $q$ is an arbitrary complex number. It has over it a (Wess-Zumino)
calculus $\Omega^*({\cal A})$ \cite{PuszWoronowicz,WZ} which is 
invariant under the action of a quantum group.  An arbitrary element
$\xi$ of $\Omega^1({\cal A})$ can be written uniquely in the form 
$\xi = \xi_x dx + \xi_y dy$ where $\xi_x$ and $\xi_y$ are elements of
${\cal A}$ but the basis elements $dx$ and $dy$ do not anti-commute and
they do not commute with the elements of the algebra.  There is no
generalized Dirac operator.

\subsection{The extended quantum plane}

One can extend the previous algebra by adding the inverses $x^{-1}$ and
$y^{-1}$. For each integer $n$ and each set of $n$ linear-independent
elements $\lambda_i$ of ${\cal A}$, there exists then a differential 
calculus $\Omega^*({\cal A})$ \cite{DimakisMadore,MadoreMouradc} based 
on the derivations
\begin{equation}
e_i f = [\lambda_i, f].
\end{equation}
The differential of an element of $f$ of ${\cal A}$ is defined as in the
usual case by the identity
\begin{equation}
df(e_i) = e_i f.
\end{equation} 
The $\Omega^1({\cal A})$ has a preferred basis of $n$ elements
$\theta^i$ which commute with the elements of the algebra.  The
relations of the calculus can be written in the form
\begin{equation}
\theta^i \theta^j = P^{ij}{}_{kl} \theta^k \theta^l
\end{equation}
where the $P^{ij}{}_{kl}$ are functions of $q$. In this case also the
special 1-form $\theta = - \lambda_i \theta^i$ is a generalized Dirac
operator.  If one accepts as definition of `dimension' the rank of
$\Omega^1({\cal A})$ then for $q \neq 1$ the `dimension' of the extended
quantum plane can be arbitrary. Unless however $n=2$ the differential
calculus has a singular limit as $q \rightarrow 1$. For $n=2$ and a
special choice of the elements $\lambda_i$ the resulting differential
calculus is an extension of the Wess-Zumino calculus.  The $\theta$
cannot however be considered as an element of the Wess-Zumino calculus
since the $\theta^i$ are constructed using the inverses of $x$ and $y$.

\subsection{Quantum groups}

Consider the quantum groups $GL_q(n)$ with generators $T^i_j$
and antipode $\kappa$.  The left invariant 1-forms
$$
\omega^i_j\ = \kappa(T^i_k) dT^k_j
$$
generate \cite{Woronowicz} a bicovariant
differential calculus $\Omega^*\big(GL_q(n)\big)$.
There is a right- and left-invariant generalized Dirac operator.

\section{Kaluza-Klein Theory}

One can consider a modified version of Kaluza-Klein with a
noncommutative algebra to describe the extra hidden `dimensions' and
study electromagnetism or gravity on the extended structure. There have
been numerous models based on this idea.  With electromagnetism one can
obtain probably any Yang-Mills-Higgs-Kibble model by appropriately
chosing the noncommutative factor of the differential calculus. The
Higgs potential appears as the term of the electromagnetic action
associated with the hidden `directions'.  The length scale of the extra
texture which space-time aquires must be of the order of the weak-boson
Compton wave length.

With gravity the situation is more rigid. No matter what the algebra and
the differential calculus, Yang-Mills potentials can be only associated
with those `directions' in the hidden `dimensions' for which the 1-forms
commute with all the elements of the algebra. The length scale must be
here of the order of the Planck length.  We refer, for example, to a
recent review article \cite{MadoreMouradb} for further details and
references to the original literature.

\section{Classical Gravity}

Classical gravity can be introduced in different ways. We shall define
it in a way which seems best suited to a noncommutative generalization.
The algebra ${\cal A}$ is the algebra of smooth functions on the
space-time manifold.  As principal steps we mention the following
possible list:
\begin{trivlist}
\item[1.] Introduce a moving frame
(Vierbein) $\theta^\alpha$ which is a basis of $\Omega^1({\cal A})$. We
here suppose for simplicity that the manifold is parallelizable. 
\item[2.] Define a metric using the frame
$$
g(\theta^\alpha \otimes \theta^\beta) = g^{\alpha\beta}
$$
where the $g^{\alpha\beta}$ are given elements of ${\cal A}$.
\item[3.] Define a covariant derivative on the frame
\begin{equation}
D\theta^\alpha = 
- \omega^\alpha{}_{\beta\gamma}\theta^\beta \otimes \theta^\gamma
\label{covder}
\end{equation}
where the $\omega^\alpha{}_{\beta\gamma}$ are given elements of 
${\cal A}$ and extend to arbitrary 1-forms 
$\xi = \xi_\alpha \theta^\alpha$ using the Leibniz rule
$$
D (\xi_\alpha \theta^\alpha) =  d\xi_\alpha \otimes \theta^\alpha + 
\xi_\alpha D\theta^\alpha.
$$
\item[4.] Require that the torsion vanish.
\item[5.] Require that the connection be metric.
\item[6.] Define the curvature in terms of the second covariant
derivative $D^2$ and contract indices to form the Ricci scalar.
\item[7.] Use the integral over the manifold to define the
Einstein-Hilbert action,
\end{trivlist}

\section{Linear Connections}

In the noncommutative case we try as much as possible to mimic the
steps of the classical case. Let ${\cal A}$ be an arbitrary algebra 
and $\Omega^*({\cal A})$ a differential calculus over it. If 
$\Omega^1({\cal A})$ has a basis $\theta^\alpha$ (Stehbein) then one can
repeat the first 5 steps above, with a modification of Step 3. We have
already mentioned that there are problems with Steps 6. and 7.
Since some of the interesting models use differential calculi which do
not possess a frame we discuss covariant derivatives in general. We replace
Step 3. by
\begin{trivlist}
\item[$3^\prime.$] Define a covariant derivative as a linear map of the
form
\begin{equation}
\Omega^1({\cal A}) \buildrel D \over \rightarrow 
\Omega^1({\cal A}) \otimes_{\cal A} \Omega^1({\cal A}).
\end{equation}
This is (\ref{covder}) in a slightly more abstract notation.  The
subscript on the tensor product means that for an arbitrary element $f$
of ${\cal A}$ and elements $\xi$ and $\eta$ of $\Omega^1({\cal A})$ one
has the identity 
$$
\xi f \otimes \eta = \xi \otimes f \eta. 
$$
The Leibniz rule must be extended since in general
$$
f \xi \neq  \xi f.
$$
\end{trivlist}

\subsection{The Leibniz rules}

Since nothing commutes in general the covariant derivative must satisfy
a left Leibniz rule
\begin{equation}
D (f \xi) =  df \otimes \xi + f D\xi,
\end{equation}
as well as a right Leibniz rule
\begin{equation}
D(\xi f) = \sigma (\xi \otimes df) + (D\xi) f.
\end{equation}
The purpose of the map $\sigma$ \cite{Mourad} is to bring the
differential to the left while respecting the order of the factors. In
the commutative case it is a simple permutation. On the Wess-Zumino
calculus over the quantum plane \cite{DuboisVioletteM3b} and on the
Woronowicz calculus over quantum groups \cite{GeorgelinM3} it is given 
by the $R$-matrix. It is necessarily \cite{DuboisVioletteM3b} bilinear,
$$
\sigma (f \xi \otimes \eta) = f \sigma (\xi \otimes \eta),\quad
\sigma (\xi \otimes \eta f) = \sigma (\xi \otimes \eta) f
$$
and it must satisfy a compatibility condition \cite{DuboisVioletteM3b}.
As definition of a linear connection in the noncommutative case we
propose the couple $(D, \sigma)$.

\section{Gravity on Fuzzy Spaces}

The definition we propose of gravity on a fuzzy space-time is a
simple-minded generalization of the classical case. The only
modification is the introduction of the generalized permutation
$\sigma$.  We first discuss the models and then an example of fuzzy
gravity.  When there is a generalized Dirac operator $\theta$ it is easy
to see that at least one covariant derivative can be defined by the
formula
\begin{equation}
D\xi = - \theta \otimes \xi + \sigma(\xi \otimes \theta).       \label{Dxi}
\end{equation}
When $\sigma = -1$ there is a simple relation between the corresponding
covariant derivative and the exterior derivative $d$.

\subsection{The models}

On the first Connes-Lott model there is no covariant derivative with
$D^2 \neq 0$. On the second model there is a 1-parameter family of
$\sigma$ and for each $\sigma$ there is a unique covariant derivative,
given by (\ref{Dxi}), which is torsion free and, for one value of the
parameter, compatible with a metric \cite{M3}.

On the matrix models with derivation-based differential calculi there is
a natural metric defined by requiring that the frame be orthonormal.
There is a natural $\sigma$, defined as a permutation on the frame,
with respect to which there is a unique torsion-free, metric-compatible
covariant derivative \cite{DuboisVioletteKernerMadore,M3}. The covariant
derivative (\ref{Dxi}) has torsion but no curvature.

On the quantum plane with generic $q$ and with the Wess-Zumino calculus
the unique $\sigma$ is given in terms of the $R$-matrix. There is, to
within a normalization constant $\mu$, a unique covariant derivative. It
is without torsion but not metric-compatible. In the limit 
$q \rightarrow 1$ the covariant derivative yields as curvature a quadratic
polynomial in $x$ and $y$
\cite{DuboisVioletteM3b}:
\begin{equation}
R^i{}_{j12} = 4 \mu^4 A^i_j, \qquad 
A = \left(\begin{array}{cc}
xy &-x^2\\ y^2 &-xy
\end{array}\right).
\end{equation}

On the extended quantum plane for generic $q$ and for each differential
calculus there is a unique torsion-free, metric connection which yields
a Gaussian curvature on the plane in the limit $q \rightarrow 1$.  In
particular the Wess-Zumino calculus is associated to the Gaussian
curvature
\begin{equation}
K = x^{-4} (1 + y^4).
\end{equation}
This curvature remains as a shadow of the Wess-Zumino calculus.

On the quantum groups $GL_q(n)$ for each $\sigma$ the only linear
connection for generic $q$ is the one defined by (\ref{Dxi}). The
arbitrariness lies alone in the generalized permutation $\sigma$ for
which there exists at least a 2-parameter family \cite{GeorgelinM3}.

\subsection{Fuzzy gravity}

In defining gravity on a fuzzy space-time we follow \cite{MadoreMouradc}
closely the example of the extended quantum plane \cite{DimakisMadore}.
Consider a $*$-algebra ${\cal A}$ with 4 hermitian generators $q^\mu$ 
which satisfy (\ref{comrel}) as well as the condition \cite{DFR}
\begin{equation}
[q^\lambda, q^{\mu\nu}] = 0.                                 \label{comrel2}
\end{equation}
We shall suppose that the inverse $q^{-1}_{\mu\nu}$ of $q^{\mu\nu}$ exists:
$q^{-1}_{\lambda\mu}q^{\mu\nu} = \delta^\nu_\lambda$.

Consider the derivations $e_\lambda$ of ${\cal A}$ given by
\begin{equation}
e_\lambda f = - i \mu_P^2 q^{-1}_{\lambda\mu} [q^\mu, f]
\end{equation}
and define a differential as usual by the equation 
$df(e_\lambda) = e_\lambda f$. The frame is given by
\begin{equation}
\theta^\lambda = dq^\lambda
\end{equation}
The unique covariant derivative is given by $D\theta^\lambda = 0$.
That is, the algebra is a noncommutative version of Minkowski space,
extended by the `coordinates' $q^{\mu\nu}$.

If one perturbs the relation (\ref{comrel2}) by
\begin{equation}
q^{\mu\nu} \mapsto q^{\prime\mu\nu} = q^{\mu\nu} + q^{\mu\nu}_{(1)}
\end{equation}
with $[q^\lambda, q^{\mu\nu}_{(1)}]$ in the center of the algebra
then there is a rather uninteresting but not completely trivial
covariant derivative given by (\ref{covder}) with
$\omega^\lambda{}_{\mu\nu}$ given in terms of 
$q^{-1}_{\mu\rho}q^{-1}_{\nu\sigma}[q^\lambda, q^{\rho\sigma}_{(1)}]$.
A preliminary investigation has been made \cite{MadoreMouradc} of
fuzzy de~Sitter space.

\end{document}